\documentclass[aps,pra,reprint,showpacs,superscriptaddress,twocolumn]{revtex4-1}%
\usepackage{amssymb}
\usepackage{mathrsfs}
\usepackage{amsfonts}
\usepackage{graphicx}
\usepackage{amsmath}
\usepackage{dcolumn}
\usepackage{color}
\usepackage[colorlinks,linkcolor=blue,citecolor=blue,hyperindex,bookmarks=false,pdfstartview=FitH]{hyperref}
\begin{document}
\title{Controllable optical response and tunable sensing based on self interference in waveguide QED systems}
\author{Lei Du}
\affiliation{Beijing Computational Science Research Center, Beijing 100193, China}
\author{Zhihai Wang}
\affiliation{Center for Quantum Sciences and School of Physics, Northeast Normal University, Changchun 130024, China}
\author{Yong Li}
\email{liyong@csrc.ac.cn}
\affiliation{Beijing Computational Science Research Center, Beijing 100193, China}
\affiliation{Center for Quantum Sciences and School of Physics, Northeast Normal University, Changchun 130024, China}
\affiliation{Synergetic Innovation Center for Quantum Effects and Applications, Hunan Normal University, Changsha 410081, China}

\date{\today }

\begin{abstract}
We study the self interference effect of a resonator coupled with a bent waveguide at two separated ports. Such interference effects are shown to be similar for the cases of standing-wave and traveling-wave resonators, while in the system of two separated resonators indirectly coupled via a waveguide, the coupling forms and the related interference effects depend on which kind of resonators is chosen. Due to the self interference, controllable optical responses including tunable linewidth and frequency shift, and optical dark state can be achieved. Moreover, we consider a self-interference photon-magnon hybrid model and show phase-dependent Fano-like line shapes which have potential applications in frequency sensing. The photon-magnon hybridization can not only enhance the sensitivity and provide tunable working region, but also enables optical readout of the magnetic field strength in turn. The results in this paper provide a deeper insight into the self interference effect and its potential applications.
\end{abstract}
\maketitle

\section{Introduction}
Waveguide quantum electrodynamics (QED), where emitters are coupled to a continuum of traveling photons confined in one-dimensional open waveguides, provides a promising platform for enhancing light-matter interactions and suppressing dissipations into the surrounding environment~\cite{wQED1,wQED2}. This field has been sufficiently explored with various candidates such as superconducting circuits~\cite{circuitS1,circuitS2,circuitNP}, optical waveguides~\cite{optgd1,optgd2}, and coupled-resonator arrays~\cite{array1,array2}. Up to now, a series of novel phenomena such as chiral photon-atom interactions~\cite{chiralcp1,chiralcp2}, phase transitions~\cite{pt1,pt2}, topologically induced unconventional quantum optics~\cite{tuqo}, and single-photon nonreciprocity~\cite{nr1,nr2,nr3} have been achieved based on various techniques and engineered configurations. In particular, waveguide QED is theoretically predicted~\cite{fan1999,IEEEo,xiao2008,xiao2010,jhli} and experimentally verified~\cite{WongPRL,fan2010,BBLi} to enable indirect couplings between spatially separated emitters, which have potential applications in large-scale quantum network. 

On the other hand, cavity magnonics based on photon-magnon hybridization provides an excellent solid platform for quantum information processing~\cite{cm1,cm2,cm3,cm4,cm5,cm6}. Over the past few years, it has attracted much attention due to the ability to achieve strong couplings in microwave regime and has incited a lot of breakthroughs, for example, gradient memory~\cite{gradient}, logic gate~\cite{logic}, magnon blockade~\cite{mblock}, and photon-magnon-photon coupling (cavity magnomechanics)~\cite{magSATang}, just to name a few. Recently, the idea of waveguide QED is introduced to cavity magnonics to achieve level attraction~\cite{LAmag} and giant nonreciprocity~\cite{magnr}. Along this line, an indirect coupling scheme for remote cavity and magnon modes is proposed~\cite{magref}, where the effective coupling can be purely dissipative by tuning the separation distance between them. 

In this paper, we begin with revisiting the well-studied model where two separated modes are indirectly coupled via a common waveguide and then generalize it to a single-resonator model. This model is formed by coupling a resonator with a bent waveguide at two different ports such that photons can travel via either external waveguide or intra-resonator path from one port to another, leading to the self interference effect. It shows that the phase factor induced by the separation between the two ports plays a key role which modifies the resonance frequency, linewidth, and the input term simultaneously. In particular, an optical dark state can be tailored judiciously with which the resonator is effectively decoupled from the waveguide. Moreover, we propose a sensing scheme based on a photon-magnon hybrid model where the self-interference resonator mentioned above is coupled with a ferromagnetic material via the magnetic dipole-dipole interaction. Although self interference has been used for both dispersive and dissipative sensing methods in bare-resonator systems (only an empty resonator is coupled to the waveguide)~\cite{disper1,disper2,disper3,PRDong,sensJOSAB}, the hybrid model here shows a series of advantages. On one hand, the sensitivity can be markedly improved due to the hybridization induced sharp Fano-like line shapes. The sensing performance can be further optimized by tuning the phase factor and the optimal working region can be changed due to the tunable resonance frequency of the magnon mode. On the other hand, the hybridization in turn enables sensing for the strength of the magnetic field, implying that our scheme can be used as a high-performance   magnetometer~\cite{mapp1,mapp2,mapp3,mmeter1,mmeter2,mmeter3,mmeter4}.

\section{Model and equations}\label{sec1}

\begin{figure}[ptb]
\centering
\includegraphics[width=7 cm]{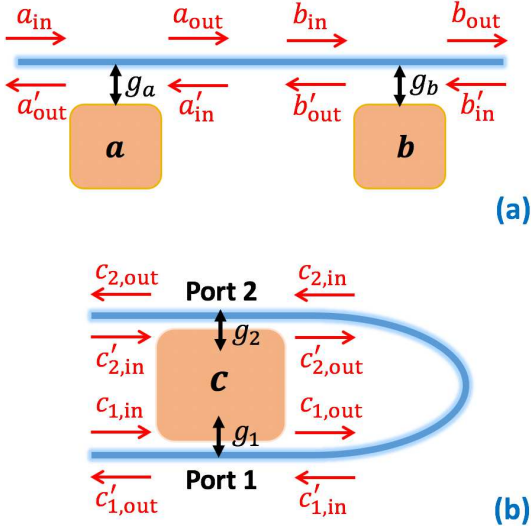}
\caption{Schematic diagrams of (a) two separated standing-wave resonators side-coupled with a straight waveguide and (b) a single standing-wave resonator side-coupled with a bent waveguide at two separated ports.}\label{fig1}
\end{figure}

We first revisit a general model in which indirect couplings between remote modes can be achieved. As shown in Fig.~\ref{fig1}(a), two spatially separated standing-wave resonators $a$ and $b$ are side-coupled to a common waveguide. The separation distance $L$ is much larger than the wavelengths of the intra-resonator fields, thus there is no direct coupling between the two resonators due to the absence of modal overlap. On the other hand, $L$ is assumed to be much smaller than the coherence length of photons in the waveguide to avoid obvious non-Markovian retarded effects~\cite{Solano,longhiret1,longhiret2}. The Hamiltonian can be written as ($\hbar=1$)~\cite{xiao2008,xiao2010,walls}  
\begin{equation}
H=\sum_{j=a,b}\omega_{j}j^{\dag}j+i\sqrt{\kappa_{a,e}}(j_{\textrm{in}}+j_{\textrm{in}}')(j^{\dag}-j),
\label{eq1}
\end{equation}
which leads to the dynamic equations
\begin{equation}
\begin{split}
&\frac{da}{dt}=-(i\omega_{a}+\kappa_{a})a+\sqrt{\kappa_{a,e}}(a_{\textrm{in}}+a_{\textrm{in}}'),\\
&\frac{db}{dt}=-(i\omega_{b}+\kappa_{b})b+\sqrt{\kappa_{b,e}}(b_{\textrm{in}}+b_{\textrm{in}}').
\label{eq2}
\end{split}
\end{equation}
Here $\omega_{j}$ is the resonance frequency of resonator $j$. $\kappa_{j}=\kappa_{j,i}+\kappa_{j,e}$ is the total loss of resonator $j$, with $\kappa_{j,i}$ the intrinsic loss due to photon leakage and $\kappa_{j,e}$ the external loss due to the coupling with the waveguide. Note that the external losses have been doubled here because both resonators are assumed to be two-sided with mirror symmetry~\cite{walls}. Experimentally, the external loss rate is determined by the resonator-waveguide coupling strength, i.e., $\kappa_{j,e}=2\pi g_{j}^{2}$ with $g_{j}$ the (real) coupling strength between resonator $j$ and the waveguide~\cite{xiao2010,jhli,Solano}. As shown in Fig.~\ref{fig1}(a), $a_{\textrm{in}}$ and $a_{\textrm{in}}'$ ($b_{\textrm{in}}$ and $b_{\textrm{in}}'$) are the input fields coming from the left and right sides of resonator $a$ ($b$), respectively. Similarly, we define $a_{\textrm{out}}$ and $a_{\textrm{out}}'$ ($b_{\textrm{out}}$ and $b_{\textrm{out}}'$) as respectively the output fields leaving from the right and left sides of resonator $a$ ($b$). In this case, the input and output fields obey the following relations~\cite{walls,xiao2010}
\begin{equation}
\begin{split}
&a_{\textrm{out}}=a_{\textrm{in}}-\sqrt{\kappa_{a,e}}a,\,a_{\textrm{out}}'=a_{\textrm{in}}'-\sqrt{\kappa_{a,e}}a,\\
&b_{\textrm{in}}=a_{\textrm{out}}e^{i\phi},\,a_{\textrm{in}}'=b_{\textrm{out}}'e^{i\phi},\\
&b_{\textrm{out}}'=b_{\textrm{in}}'-\sqrt{\kappa_{b,e}}b,\,b_{\textrm{out}}=b_{\textrm{in}}-\sqrt{\kappa_{b,e}}b,
\end{split}
\label{eq3}
\end{equation}
where $\phi$ is the phase accumulated by photons traveling in the waveguide from one resonator to another, determined by both the separation distance between the two resonators and the wave vector of photons traveling in the waveguide~\cite{fan1999,IEEEo,xiao2008,xiao2010,jhli,Solano,longhiret1,longhiret2}. Substituting Eq.~(\ref{eq3}) to Eq.~(\ref{eq2}), we can obtain
\begin{equation}
\begin{split}
&\frac{da}{dt}=-(i\omega_{a}+\kappa_{a})a-\sqrt{\kappa_{a,e}\kappa_{b,e}}e^{i\phi}b+\sqrt{\kappa_{a,e}}f_{a,\textrm{in}},\\
&\frac{db}{dt}=-(i\omega_{b}+\kappa_{b})b-\sqrt{\kappa_{a,e}\kappa_{b,e}}e^{i\phi}a+\sqrt{\kappa_{b,e}}f_{b,\textrm{in}},
\label{eq4}
\end{split}
\end{equation}
where $f_{a,\textrm{in}}=a_{\textrm{in}}+b_{\textrm{in}}'e^{i\phi}$ and $f_{b,\textrm{in}}=b_{\textrm{in}}'+a_{\textrm{in}}e^{i\phi}$. Eq.~(\ref{eq4}) shows that indirect coupling between the two separated modes can be achieved via traveling photons in the waveguide. Such an indirect coupling is non-Hermitian due to the identical phase accumulation $\phi$ for both directions [see from the identical coefficient $-\sqrt{\kappa_{a,e}\kappa_{b,e}}e^{i\phi}$ in both equations of Eq.~(\ref{eq4})]. In particular, purely dissipative couplings can be achieved when $\phi=n\pi$ ($n$ is an arbitrary integer). 

Now we generalize the theory above by considering a single standing-wave resonator $c$ coupled with a bent waveguide at two separated ports, as shown in Fig.~\ref{fig1}(b). In this case, the dynamic equation of $c$ is given by
\begin{equation}
\begin{split}
\frac{dc}{dt}=&-(i\omega_{c}+\kappa_{c})c+\sqrt{\kappa_{1,e}}(c_{\textrm{in},1}+c_{\textrm{in},1}')\\
&+\sqrt{\kappa_{2,e}}(c_{\textrm{in},2}+c_{\textrm{in},2}'),
\end{split}
\label{eq5}
\end{equation}
where $\omega_{c}$ is the resonance frequency of resonator $c$. $\kappa_{c}=\kappa_{c,i}+\kappa_{1,e}+\kappa_{2,e}$ is the total loss of $c$, with $\kappa_{c,i}$ the intrinsic loss and $\kappa_{1(2),e}$ the external loss at port $1$ ($2$). $\kappa_{1(2),e}=2\pi g_{1(2)}^{2}$ depends on the resonator-waveguide coupling strength $g_{1(2)}$ at port $1$ ($2$). Note that in this case, the input field of resonator $c$ contains four parts: $c_{j,\textrm{in}}$ and $c_{j,\textrm{in}}'$ (corresponding to the input parts coming from the left and right sides of port $j$, respectively) with $j=1,\,2$. Similarly, the output field contains four parts $c_{j,\textrm{out}}$ and $c_{j,\textrm{out}}'$, as shown in Fig.~\ref{fig1}(b). In this case, we have
\begin{equation}
\begin{split}
&c_{1,\textrm{out}}=c_{1,\textrm{in}}-\sqrt{\kappa_{1,e}}c,\,c_{1,\textrm{out}}'=c_{1,\textrm{in}}'-\sqrt{\kappa_{1,e}}c,\\
&c_{2,\textrm{in}}=c_{1,\textrm{out}}e^{i\phi},\,c_{1,\textrm{in}}'=c_{2,\textrm{out}}'e^{i\phi},\\
&c_{2,\textrm{out}}'=c_{2,\textrm{in}}'-\sqrt{\kappa_{2,e}}c,\,c_{2,\textrm{out}}=c_{2,\textrm{in}}-\sqrt{\kappa_{2,e}}c.
\end{split}
\label{eq6}
\end{equation}
With Eqs.~(\ref{eq5}) and (\ref{eq6}), we can obtain the effective dynamic equation of $c$
\begin{equation}
\begin{split}
\frac{dc}{dt}=&-(i\omega_{c}+\kappa_{c})c-2\sqrt{\kappa_{1,e}\kappa_{2,e}}e^{i\phi}c\\
&+\sqrt{\kappa_{1,e}}f_{1,\textrm{in}}+\sqrt{\kappa_{2,e}}f_{2,\textrm{in}},
\end{split}
\label{eq7}
\end{equation}
where $f_{1,\textrm{in}}=c_{1,\textrm{in}}+c_{2,\textrm{in}}'e^{i\phi}$ and $f_{2,\textrm{in}}=c_{2,\textrm{in}}'+c_{1,\textrm{in}}e^{i\phi}$. One can also obtain from Eq.~(\ref{eq6}) the total input-output relations of this model as
\begin{equation}
\begin{split}
&c_{2,\textrm{out}}=(c_{1,\textrm{in}}-\sqrt{\kappa_{1,e}}c)e^{i\phi}-\sqrt{\kappa_{2,e}}c,\\
&c_{1,\textrm{out}}'=(c_{2,\textrm{in}}'-\sqrt{\kappa_{2,e}}c)e^{i\phi}-\sqrt{\kappa_{1,e}}c.
\end{split}
\label{eq8}
\end{equation}
In the case of $\kappa_{1,e}=\kappa_{2,e}=\kappa_{e}$ (i.e., $g_{1}=g_{2}$), Eqs.~(\ref{eq7}) and (\ref{eq8}) can be simplified as
\begin{equation}
\begin{split}
&\frac{dc}{dt}=-[i\omega_{c}+\kappa_{c,i}+2\kappa_{e}(1+e^{i\phi})]c+\sqrt{\kappa_{e}}(1+e^{i\phi})c_{1,\textrm{in}},\\
&c_{2,\textrm{out}}=c_{1,\textrm{in}}e^{i\phi}-\sqrt{\kappa_{e}}(1+e^{i\phi})c,\\
&c_{1,\textrm{out}}'=c_{2,\textrm{in}}'e^{i\phi}-\sqrt{\kappa_{e}}(1+e^{i\phi})c.
\end{split}
\label{eq9}
\end{equation}

Equations~(\ref{eq7})-(\ref{eq9}) demonstrate a self interference effect which may significantly modify the optical properties of the system. We will discuss this effect in detail in the next two sections. As a supplement, we provide in Appendix~\ref{appA} an alternative method based on real-space Schr\"{o}dinger equation to verify our conclusion at single-photon level, which leads to essentially the same results. Moreover, we point out that for the case of Fig.~\ref{fig1}(a), the effective dynamic equations can be quite different if $a$ and $b$ are traveling-wave resonators, while for the case of Fig.~\ref{fig1}(b), the self interference effects are shown to be similar whether $a$ and $b$ are standing-wave or traveling-wave resonators. The related details are shown in Appendix~\ref{appB}.

\section{Phase-dependent optical response}\label{sec2}

\begin{figure}[ptb]
\centering
\includegraphics[width=8.5 cm]{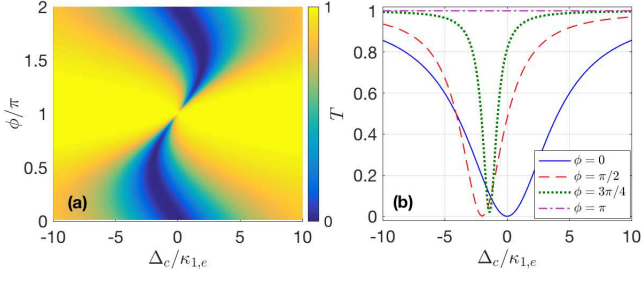}
\caption{(a) Pseudo-color map of transmission rate $T$ versus detuning $\Delta_{c}$ and phase $\phi$. (b) Profiles of $T$ versus $\Delta_{c}$ with different values of $\phi$. The other parameters are $\kappa_{c,i}/\kappa_{1,e}=0.1$ and $\kappa_{2,e}/\kappa_{1,e}=1$.}\label{fig2}
\end{figure} 

To study how the separation between the two ports affects the optical response of the single-resonator model in Fig.~\ref{fig1}(b), we consider an external input signal injected from the lower side of the waveguide. In this case, $c_{1,\textrm{in}}\rightarrow c_{1,\textrm{in}}+\varepsilon_{s}e^{-i\omega_{s}t}$ comprises both the vacuum input field and external signal, with $\varepsilon_{s}$ and $\omega_{s}$ the amplitude and frequency of the signal, respectively. Under the rotating frame with respect to $\omega_{s}$, the effective dynamic equation of the mean value of $c$ can be written as  
\begin{equation}
\begin{split}
\frac{d\langle c\rangle}{dt}=&-(i\Delta_{c}+\kappa_{c}+2\sqrt{\kappa_{1,e}\kappa_{2,e}}e^{i\phi})\langle c\rangle\\
&+(\sqrt{\kappa_{1,e}}+\sqrt{\kappa_{2,e}}e^{i\phi})\varepsilon_{s},
\end{split}
\label{eq10}
\end{equation}
where $\Delta_{c}=\omega_{c}-\omega_{s}$ is the detuning between resonator $c$ and the input signal. Here the vacuum input terms have been dropped due to their zero mean values. By solving the steady-state solution of Eq.~(\ref{eq10}) which reads
\begin{equation}
c_{s}=\frac{(\sqrt{\kappa_{1,e}}+\sqrt{\kappa_{2,e}}e^{i\phi})\varepsilon_{s}}{i\Delta_{c}+\kappa_{c}+2\sqrt{\kappa_{1,e}\kappa_{2,e}}e^{i\phi}},
\label{eq11}
\end{equation} 
one can study the mean response of the model. In this case, the steady-state output field at the upper side of the waveguide can be given by
\begin{equation}
c_{2,\textrm{out}}^{s}=(c_{1,\textrm{in}}-\sqrt{\kappa_{1,e}}c_{s})e^{i\phi}-\sqrt{\kappa_{2,e}}c_{s}
\label{eq12}
\end{equation}
with which we can define the transmission rate of the input signal $T=|c_{2,\textrm{out}}^{s}/\varepsilon_{s}|^{2}$. It is worth noting that the transmission rate in our side-coupled case (the resonator is side-coupled with the waveguide) equals exactly to the reflection rate in the directly-coupled case (the resonator is directly placed into the waveguide). This is because in the latter case, photons in the waveguide can be transmitted only by tunneling in and out the resonator~\cite{fan2009}. 

\begin{figure}[ptb]
\centering
\includegraphics[width=8.5 cm]{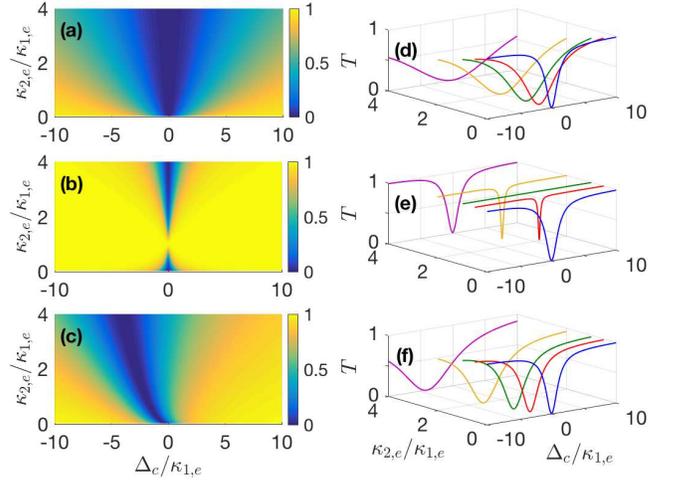}
\caption{Pseudo-color maps of transmission rate $T$ versus detuning $\Delta_{c}$ and external decay rate $\kappa_{2,e}$ with (a) $\phi=0$, (b) $\phi=\pi$, (c) $\phi=\pi/2$. Profiles of $T$ versus $\Delta_{c}$ with different values of $\kappa_{2,e}$ and (d) $\phi=0$, (e) $\phi=\pi$, (f) $\phi=\pi/2$. The other parameter is $\kappa_{c,i}/\kappa_{1,e}=0.1$.}\label{fig3}
\end{figure}

Figure \ref{fig2}(a) shows the transmission rate $T$ versus the detuning $\Delta_{c}$ and phase $\phi$. It is clear that the absorption window can be significantly modified by tuning the phase $\phi$ (i.e., the separation distance between the two ports) and the dependence of $T$ on $\phi$ is $2\pi$-periodic. As $\phi$ increases from $0$ to $\pi$ [or from $2n\pi$ to $(2n+1)\pi$ with $n$ an arbitrary integer], the width of the transmission dip decreases gradually, while the position of the dip shows a non-monotonic behavior, i.e., it first moves towards left (the direction towards negative values) and then returns back to the resonance position $\Delta_{c}=0$. Due to the relatively small intrinsic loss, the transmission dip implies in fact a reflection enhancement rather than a strong resonant absorption. To show the details clearly, we plot in Fig.~\ref{fig2}(b) the profiles of $T$ versus $\Delta_{c}$ with different values of $\phi$. Indeed, one can find that the transmission dip becomes narrower and narrower as $\phi$ increases from $0$ and its position shift reaches the maximum near $\phi=\pi/2$. This can be understood by the term $-2\sqrt{\kappa_{1,e}\kappa_{2,e}}e^{i\phi}$ in Eq.~(\ref{eq10}), which is complex for $\phi\neq n\pi$. The real part corresponds to a modification of the linewidth, while the imaginary part corresponds to a frequency shift. Clearly, the linewidth (frequency shift) of resonator $c$ reaches the minimum (maximum) at $\phi=(2n+1)\pi$ [$\phi=(n+1/2)\pi$]. In particular, the transmission dip disappears completely when $\phi=\pi$, as shown in Fig.~\ref{fig2}(b). In this case, the resonator cannot be excited by (or decay to) the waveguide, which demonstrates an optical dark state~\cite{synphonon}. One can also understand this result by the term $(\sqrt{\kappa_{1,e}}+\sqrt{\kappa_{2,e}}e^{i\phi})\varepsilon_{s}$ in Eq.~(\ref{eq10}), which shows that the two input parts completely cancel each other when $\kappa_{1,e}=\kappa_{2,e}$ and $\phi=(2n+1)\pi$.   

It is worth pointing out that the external decay rates also play an important role in controlling the optical response, which can be tuned experimentally by changing the distance between the resonator and the waveguide. In view of this, we plot in Figs.~\ref{fig3}(a)-\ref{fig3}(c) the transmission rate $T$ versus $\Delta_{c}$ and $\kappa_{2,e}$ for $\phi=0$, $\phi=\pi$, and $\phi=\pi/2$, respectively, and plot in Figs.~\ref{fig3}(d)-\ref{fig3}(f) the corresponding profiles of $T$ with a set of chosen $\kappa_{2,e}$ to show more details. The transmission rate shows quite different dependence on $\kappa_{2,e}$ for the three cases. In the case of $\phi=0$, the width of the absorption window is positively associated with $\kappa_{2,e}$, while the depth of the window maintains almost invariant, i.e., the increasing $\kappa_{2,e}$ almost does not reduce the reflection rate. In the case of $\phi=\pi$, however, both width and depth of the window strongly depends on $\kappa_{2,e}$. As $\kappa_{2,e}$ increases from $0$ to the value of $\kappa_{1,e}$, the window narrows and shallows rapidly until it disappears completely at $\phi=\pi$ as discussed above. Further increasing $\kappa_{2,e}$ leads to an inverse but much slower process, i.e., the window becomes wider and deeper gradually as $\kappa_{2,e}$ increases. When $\kappa_{2,e}/\kappa_{1,e}=4$, the transmission profile becomes almost identical as that for $\kappa_{2,e}=0$. For the more general case of $\phi\neq n\pi$, such as the case of $\phi=\pi/2$ in Figs.~\ref{fig3}(c) and \ref{fig3}(f), the increasing $\kappa_{2,e}$ yields a wider transmission dip while the depth of the dip is insensitive to $\kappa_{2,e}$. This is a bit similar to the case of $\phi=2n\pi$. However, one can also observe a position shift that is proportional to $\kappa_{2,e}$, which is distinct from the other two cases.  

\section{Tunable sensing with photon-magnon hybridization}\label{sec3}

Now we consider that the resonator $c$ in Fig.~\ref{fig1}(b) is a microwave cavity which contains a yttrium iron garnet (YIG) sphere. The YIG sphere can be described by a uniform magnon mode $m$ (Kittel mode), which is essentially the collective motion of spins in it~\cite{cm1,cm2,cm3,cm4,cm5,magSATang}. The magnon mode shows a tunable resonance frequency $\omega_{m}$ which is determined by the external magnetic field $B$, i.e., $\omega_{m}=\gamma B$ with $\gamma$ the gyromagnetic ratio. With the rotating-wave approximation, the magnon-photon coupling can be described by $J(a^{\dag}m+m^{\dag}a)$, which originates from the magnetic dipole-dipole interaction. Using the method developed in Sec.~\ref{sec1} and assuming that an external input signal is injected from the lower side of the waveguide, the effective dynamic equations can be written as
\begin{equation}
\begin{split}
\frac{d\langle c\rangle}{dt}=&-(i\Delta_{c}+\kappa_{c}+\sqrt{\kappa_{1,e}\kappa_{2,e}}e^{i\phi})\langle c\rangle-iJ\langle m\rangle\\
&+(\sqrt{\kappa_{1,e}}+\sqrt{\kappa_{2,e}}e^{i\phi})\varepsilon_{s},\\
\frac{d\langle m\rangle}{dt}=&-(i\Delta_{m}+\kappa_{m})\langle m\rangle-iJ\langle c\rangle,
\end{split}
\label{eq13}
\end{equation}
where $\Delta_{m}=\omega_{m}-\omega_{s}$ is the detuning between magnon mode $m$ and the input signal. $\kappa_{m}$ is the magnonic decay rate. Other symbols are the same as those in Eq.~(\ref{eq10}). Once again, we can define the transmission rate
\begin{equation}
T=\Big|\frac{c_{2,\textrm{out}}}{\varepsilon_{s}}\Big|^{2}=\Big|(1-\sqrt{\kappa_{1,e}}\frac{c_{s}}{\varepsilon_{s}})e^{i\phi}-\sqrt{\kappa_{2,e}}\frac{c_{s}}{\varepsilon_{s}}\Big|^{2},
\label{eq14}
\end{equation} 
which is determined by the steady-state value of the resonator mode
\begin{equation}
c_{s}=\frac{(\sqrt{\kappa_{1,e}}+\sqrt{\kappa_{2,e}}e^{i\phi})h\varepsilon_{s}}{fh+J^{2}}
\label{eq15}
\end{equation} 
with $f=i\Delta_{c}+\kappa_{c}+2\sqrt{\kappa_{1,e}\kappa_{2,e}}e^{i\phi}$ and $h=i\Delta_{m}+\kappa_{m}$. 

\begin{figure}[ptb]
\centering
\includegraphics[width=8.5 cm]{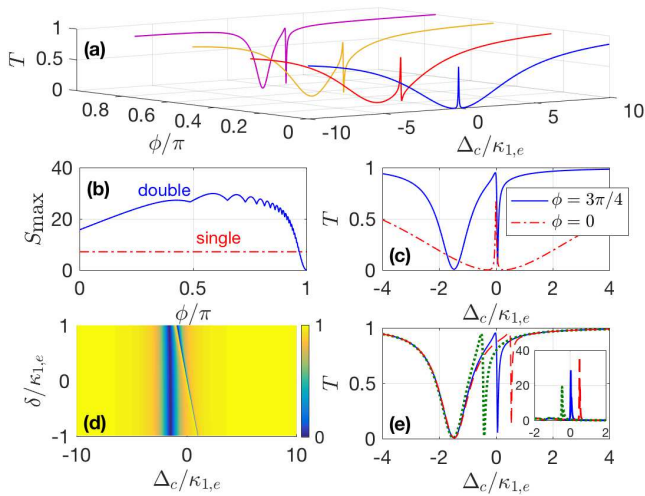}
\caption{(a) Profiles of transmission rate $T$ versus detuning $\Delta_{c}$ with $\kappa_{m}/\kappa_{c,i}=0.1$ and different values of phase $\phi$. (b) Maximal sensitivity $S_{\textrm{max}}$ versus phase $\phi$. The red dashed line shows the maximal sensitivity of the single-port model with $\kappa_{2,e}=0$ and other parameters being the same. (c) Profiles of $T$ versus $\Delta_{c}$ with $\kappa_{m}/\kappa_{c,i}=0.1$ and two chosen values of $\phi$. (d) Pseudo-color map of $T$ versus detunings $\Delta_{c}$ and $\delta$. (e) Profiles of $T$ versus $\Delta_{c}$ with different values of $\delta$. The inset in (e) shows the sensitivity $S$ versus $\Delta_{c}$ with different values of $\delta$. In panel (e) and the inset, the blue solid, red dashed and green dotted lines correspond to $\delta/\kappa_{c,i}=0$, $-0.5$ and $0.5$, respectively. Here we take $\Delta_{m}=\Delta_{c}$ in (a)-(c) and $\phi=3\pi/4$ in (d) and (e). The other parameters are $\kappa_{c,i}/\kappa_{1,e}=0.05$, $\kappa_{m}/\kappa_{1,e}=5\times10^{-3}$, $\kappa_{2,e}/\kappa_{1,e}=1$, and $J/\kappa_{1,e}=0.3$.}\label{fig4}
\end{figure}

To show the influence of the photon-magnon hybridization on the optical response, we plot in Fig.~\ref{fig4}(a) the profiles of $T$ in the case of $\Delta_{m}=\Delta_{c}$ with different values of $\phi$. Clearly, a transmission peak emerges in the presence of the YIG shpere, which splits the dip into two new ones. For $\phi=0$, the separation between the two dips equals to $2J$, which originates from the normal-mode splitting due to the magnon-photon coupling. As $\phi$ increases from $0$, the line shape becomes asymmetric (Fano-like) due to the phase induced frequency shift discussed above. During this process, the right dip narrows gradually and the transparency peak tends to be perfect ($T=1$). As a result, the transmission rate can change more drastically near the resonant point $\Delta_{c}=0$ by tuning $\phi$. Note that both dips will shallow rapidly as $\phi$ approaches $\pi$ and disappear completely for $\phi=\pi$ due to the optical dark state. For the purpose of this section, we only consider the phase far away from $(2n+1)\pi$ in the following.  

We would like to point out that the sharp Fano-like line shapes in Fig.~\ref{fig4}(a) have potential applications in sensing parameters associated with $\Delta_{c}$. Within the working region, any small perturbation of the target parameter gives rise to a drastic change in transmission rate. For the case of $\Delta_{c}=\Delta_{m}$ (i.e., $\omega_{c}=\omega_{m}$) shown in Fig.~\ref{fig4}(a), one can detect the input frequency fluctuations which are unavoidable in practice and limit the performance of dispersive sensing schemes~\cite{PRDong,noise1,noise2}. For quantitative estimation, we introduce the sensitivity $S=|dT/d\Delta_{c}|$~\cite{xiao2008} and plot in Fig.~\ref{fig4}(b) the maximal sensitivity $S_{\textrm{max}}$ over the whole frequency range as a function of $\phi$ in the case of $\Delta_{c}=\Delta_{m}$. For comparison, we also plot the maximal sensitivity of a general single-port model ($\kappa_{2,e}=0$), which is physically independent of $\phi$. It shows that the maximal sensitivity is markedly enhanced in the presence of the self interference. In addition, the performance of our sensing scheme can be further optimized by tuning the phase $\phi$, with the maximum of $S_{\textrm{max}}$ obtained at $\phi\approx0.59\pi$. Although the transmission rate is not a monotonic function of $\Delta_{c}$ near the working region, one can further determine the target parameter with different values of $\phi$. For varying $\phi$, as shown in Fig.~\ref{fig4}(c), the values of $T$ corresponding to the right side of the sharp dip changes much more drastically than those corresponding to the left side. 

One major advantage of our scheme based on cavity magnonics is the controllable frequency of the magnon mode so that the working region can be tuned by changing the strength of the external magnetic field. As shown in Fig.~\ref{fig4}(d), the position of the sharp right dip is determined by the detuning $\delta=\Delta_{c}-\Delta_{m}=\omega_{c}-\omega_{m}$ between the photon and magnon modes, while the left transmission dip is nearly unaffected by the change in $\delta$. This can be seen clearly in Fig.~\ref{fig4}(e), where we do find that the left transmission dip is quite insensitive to $\delta$. The sharp right dip, which is approximately located at $\Delta_{c}=\delta$, shows slightly changed width for different $\delta$, i.e., the width of the right dip increases mildly as $\delta$ decreases from positive to negative. The inset in Fig.~\ref{fig4}(e) depicts the sensitivity $S$ versus $\Delta_{c}$ with the three chosen values of $\delta$ in Fig.~\ref{fig4}(e). We can find that a wider transmission dip corresponds to a lower sensitivity peak. In view of this, our scheme shows better performance within the blue-detuned region ($\Delta_{c}>0$). Moreover, Figs.~\ref{fig4}(d) and \ref{fig4}(e) show that our sensing scheme maintains high performance even for $\omega_{m}\neq\omega_{c}$. Therefore our scheme can also be used for sensing fluctuations related to $\omega_{c}$ such as frequency shifts induced by the thermorefractive and thermoelastic effects~\cite{thermo1,thermo2}.

\begin{figure}[ptb]
\centering
\includegraphics[width=8.5 cm]{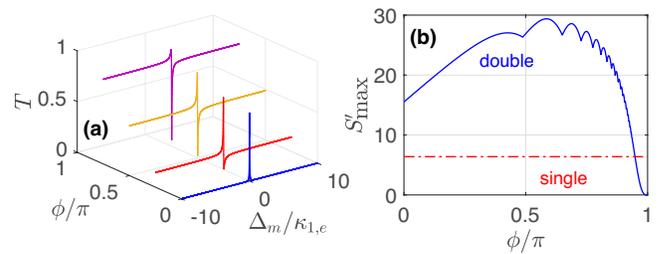}
\caption{(a) Profiles of transmission rate $T$ versus detuning $\Delta_{m}$ with different values of phase $\phi$. (b) Maximal sensitivity $S_{\textrm{max}}'$ versus phase $\phi$. The red dashed line shows the maximal sensitivity of the single-port model with $\kappa_{2,e}=0$ and other parameters being the same. The other parameters are $\Delta_{c}=0$, $\kappa_{c,i}/\kappa_{1,e}=0.05$, $\kappa_{m}/\kappa_{1,e}=5\times10^{-3}$, $\kappa_{2,e}/\kappa_{1,e}=1$, and $J/\kappa_{1,e}=0.3$.}\label{fig5}
\end{figure}

On one hand, it has been shown above that the photon-magnon hybridization can markedly enhance the sensitivity due to the sharp Fano-like line shapes and the sensitivity can be further improved by tuning the phase $\phi$. On the other hand, the transmission rate is also dependent on $\Delta_{m}$ according to Eq.~(\ref{eq14}), implying that our hybrid model can be used in turn as a solid-state magnetometer~\cite{mmeter1,mmeter2,mmeter3,mmeter4}. In other words, one can detect the resonance frequency of the magnon mode and thereby the strength of the magnetic field via the transmission rate. As shown in Fig.~\ref{fig5}(a), there is always a transmission peak near the resonant point $\Delta_{m}=\Delta_{c}=0$ (we assume $\Delta_{c}=0$ in Fig.~\ref{fig5} to show the optimal sensing performance). In the case of $\phi\neq 0$, the transmission profile becomes Fano-like similar to that in Fig.~\ref{fig4}(a), giving rise to both a peak and a dip near the resonant point. As $\phi$ increases from $0$, the transmission peak increases gradually while the depth of the transmission dip changes only slightly. Meanwhile, the off-resonance transmission rate increases markedly with $\phi$, implying that the transmission rate can change more and more drastically near the resonant point by increasing $\phi$ suitably. We plot in Fig.~\ref{fig5}(b) the maximal sensitivity $S_{\textrm{max}}'$ of our model to find out the optimal phase $\phi$ for magnetometry and that of the single-port model for comparison, with the definition of the sensitivity becoming $S'=|dT/d\Delta_{m}|$ in this case. Once again, our scheme shows much better performance than the single-port model. The $\phi$ dependence of the maximal sensitivity is quite similar with that in Fig.~\ref{fig4}(b), with the optimal performance obtained near $\phi=0.59\pi$ as well.

\section{Conclusions}

In summary, we have systematically studied the self interference effect of a resonator side-coupled with a bent waveguide at two separated ports. While the well-studied model of two indirectly coupled resonators shows quite different effective couplings for standing-wave and traveling-wave resonators, the present model supports similar self interference effect regardless of the resonator configuration. The theory is also verified at the single-photon level with the method based on real-space Schr\"{o}dinger equation. It shows that controllable optical response can be achieved by tuning the separation distance between the two ports. Besides the resonance frequency and linewidth, there is also an interference effect between input fields at the two ports, which may lead to an optical dark state under specific conditions. Moreover, the controllable Fano-like line shapes in the photon-magnon hybrid model are proved to be useful in frequency sensing. We show that the sensitivity can be markedly enhanced in the hybrid system and the working region can be changed flexibly due to the tunable resonance frequency of the magnon mode. On the other hand, our scheme can be used for magnetometry which allows for optical readout of the magnetic field strength. The results in this paper may have applications in controllable photonic transport and quantum information processing. 

\appendix
\section{Verification in real space}\label{appA}
It is known that the waveguide can be described by a bath of harmonic oscillators~\cite{walls}, therefore the total Hamiltonian of the model in Fig.~\ref{fig1}(a) can be given by
\begin{equation}
H=H_{\textrm{c}}+H_{\textrm{w}}+H_{\textrm{int}},
\label{A1}
\end{equation}
where $H_{\textrm{c}}=\omega_{a}a^{\dag}a+\omega_{b}b^{\dag}b$ and $H_{\textrm{w}}=\int\omega_{k}c_{k}^{\dag}c_{k}dk$ are the free Hamiltonians of the resonators and waveguide, respectively, and 
\begin{equation}
H_{\textrm{int}}=\int (g_{a}a^{\dag}c_{k}+g_{b}e^{ikx_{0}}b^{\dag}c_{k}+h.c.)dk,
\label{A2}
\end{equation}
is the interaction Hamiltonian between the two resonators and the waveguide under the rotating-wave approximation (RWA). The locations of resonators $a$ and $b$ are assumed to be $x=0$ and $x=x_{0}$, respectively. According to Refs.~\cite{wQED1,fan2009}, if the resonance frequencies $\omega_{a}$ and $\omega_{b}$ of both resonators are far away from the cut off frequency of the waveguide dispersion, the whole system can be described conveniently in the real space, with 
\begin{equation}
H_{\textrm{w}}=\int dx[-iv_{g}c_{R}^{\dag}(x)\frac{d}{dx}c_{R}(x)+iv_{g}c_{L}^{\dag}(x)\frac{d}{dx}c_{L}(x)]
\label{A3}
\end{equation}
and  
\begin{equation}
\begin{split}
H_{\textrm{int}}=&\int dx\{g_{a}\delta(x)[a^{\dag}c_{R}(x)+a^{\dag}c_{L}(x)+h.c.]\\
&+g_{b}\delta(x-x_{0})[b^{\dag}c_{R}(x)+b^{\dag}c_{L}(x)+h.c.]\}
\end{split}
\label{A4}
\end{equation}
in this case. Here, $v_{g}$ is the group velocity of the traveling photons in the waveguide. $c_{R}^{\dag}(x)$ [$c_{L}^{\dag}(x)$] is the bosonic operator creating a right-going (left-going) traveling photon at position $x$. Assuming that the whole system is initially prepared in the single-excitation manifold and considering that the RWA preserves the number of excitations, the wave function can be given by
\begin{equation}
\begin{split}
|\psi\rangle=&\int dx[\phi_{R}(x)c_{R}^{\dag}(x)+\phi_{L}(x)c_{L}^{\dag}(x)]|G\rangle\\
&+c_{a}a^{\dag}|G\rangle+c_{b}b^{\dag}|G\rangle,
\end{split}
\label{A5}
\end{equation} 
where $|G\rangle$ is the ground state of the whole system with no photon in the resonators and waveguide. $c_{m}$ ($m=a,\,b$) and $\phi_{j}(x)$ ($j=R,\,L$) are the excitation amplitudes of resonator $m$ and waveguide mode $c_{j}(x)$, respectively. The equations of the excitation amplitudes can be obtained by solving the Schr\"odinger equation, which reads
\begin{equation}
\begin{split}
&E\phi_{R}(x)=-iv_{g}\frac{d}{dx}\phi_{R}(x)+g_{a}\delta(x)c_{a}+g_{b}\delta(x-x_{0})c_{b},\\
&E\phi_{L}(x)=iv_{g}\frac{d}{dx}\phi_{L}(x)+g_{a}\delta(x)c_{a}+g_{b}\delta(x-x_{0})c_{e},\\
&Ec_{a}=\omega_{a} c_{a}+g_{a}[\phi_{R}(0)+\phi_{L}(0)],\\
&Ec_{b}=\omega_{b} c_{b}+g_{b}[\phi_{R}(x_{0})+\phi_{L}(x_{0})].
\end{split}
\label{A6}
\end{equation}

Now we consider that a single photon is incident from the left side of the waveguide. In this case $\phi_{R}(x)$ and $\phi_{L}(x)$ can be written as~\cite{nr3}
\begin{equation}
\begin{split}
&\phi_{R}(x)=e^{ikx}\{\theta(-x)+A[\theta(x)-\theta(x-x_{0})]+t\theta(x-x_{0})\},\\
&\phi_{L}(x)=e^{-ikx}\{r\theta(-x)+B[\theta(x)-\theta(x-x_{0})]\},
\end{split}
\label{A7}
\end{equation} 
where $\theta(x)$ is the Heaviside step function satisfying $\partial\theta(bx-a)/\partial x=b\delta(bx-a)$. Here, $A$ ($B$) denotes the amplitude of the right-going (left-going) wave in the region $0<x<x_{0}$, while $r$ ($t$) denotes the reflection (transmission) amplitude at $x=0$ ($x=x_{0}$). Substituting Eq.~(\ref{A7}) into Eq.~(\ref{A6}), one can obtain
\begin{equation}
\begin{split}
&0=-iv_{g}(A-1)+g_{a}c_{a},\\
&0=-iv_{g}(t-A)e^{ikx_{0}}+g_{b}c_{b},\\
&0=-iv_{g}(r-B)+g_{a}c_{a},\\
&0=-iv_{g}Be^{-ikx_{0}}+g_{b}c_{b},\\
&Ec_{a}=\omega_{a} c_{a}+\frac{g_{a}}{2}(A+B+1+r),\\
&Ec_{b}=\omega_{b} c_{b}+\frac{g_{b}}{2}(te^{ikx_{0}}+Ae^{ikx_{0}}+Be^{-ikx_{0}}),
\end{split}
\label{A8}
\end{equation}
where $E=v_{g}k$ for $x\neq0$ and $x\neq x_{0}$. In this way, we can obtain the effective equations of $c_{a}$ and $c_{b}$
\begin{equation}
\begin{split}
&i\frac{dc_{a}}{dt}=Ec_{a}=(\omega_{a}-i\gamma_{a})c_{a}-i\sqrt{\gamma_{a}\gamma_{b}}e^{ikx_{0}}c_{b}+g_{a},\\
&i\frac{dc_{b}}{dt}=Ec_{b}=(\omega_{b}-i\gamma_{b})c_{b}-i\sqrt{\gamma_{a}\gamma_{b}}e^{ikx_{0}}c_{a}+g_{b}e^{ikx_{0}}
\end{split}
\label{A9}
\end{equation}
with $\gamma_{a}=g_{a}^{2}/v_{g}$ and $\gamma_{b}=g_{b}^{2}/v_{g}$~\cite{wQED1}. Clearly, Eq.~(\ref{A9}) shows essentially the same effective coupling as that in Eq.~(\ref{eq4}). The input terms, however, have also the same forms if we only consider the input field coming from the left side of the waveguide ($b_{in}'=0$) in Eq.~(\ref{eq4}). 

For the model shown in Fig.~\ref{fig1}(b), Eq.~(\ref{A6}) becomes
\begin{equation}
\begin{split}
&E\phi_{R}(x)=-iv_{g}\frac{d}{dx}\phi_{R}(x)+[g_{1}\delta(x)+g_{2}\delta(x-x_{0})]c_{c},\\
&E\phi_{L}(x)=iv_{g}\frac{d}{dx}\phi_{L}(x)+[g_{1}\delta(x)+g_{2}\delta(x-x_{0})]c_{c},\\
&Ec_{c}=\omega_{c} c_{c}+g_{1}[\phi_{R}(0)+\phi_{L}(0)]+g_{2}[\phi_{R}(x_{0})+\phi_{L}(x_{0})],
\end{split}
\label{A10}
\end{equation}
where $c_{c}$ is the single-photon excitation amplitude of resonator $c$. With a similar procedure, we can obtain
\begin{equation}
\begin{split}
0=&-iv_{g}(A-1)+g_{1}c_{c},\\
0=&-iv_{g}(t-A)e^{ikx_{0}}+g_{2}c_{c},\\
0=&-iv_{g}(r-B)+g_{1}c_{c},\\
0=&-iv_{g}Be^{-ikx_{0}}+g_{2}c_{c},\\
0=&\frac{g_{2}}{2}(te^{ikx_{0}}+Ae^{ikx_{0}}+Be^{-ikx_{0}})\\
&+\frac{g_{1}}{2}(A+B+1+r).
\end{split}
\label{A11}
\end{equation}
The effective equation of $c_{c}$ can be obtained by solving Eq.~(\ref{A11}), i.e.,
\begin{equation}
\begin{split}
i\frac{dc_{c}}{dt}=Ec_{c}=&(\omega_{c}-i\gamma_{1}-i\gamma_{2})c_{c}+g_{1}+g_{2}e^{ikx_{0}}\\
&-2i\sqrt{\gamma_{1}\gamma_{2}}e^{ikx_{0}}c_{c}
\end{split}
\label{A12}
\end{equation}
with $\gamma_{1}=g_{1}^{2}/v_{g}$ and $\gamma_{2}=g_{2}^{2}/v_{g}$. Once again, the coupling and input terms in Eq.~(\ref{eq7}) (in the case of $c_{2,\textrm{in}}'=0$) have the same forms as these in Eq.~(\ref{A12}), which verifies our conclusions in Sec.~\ref{sec1}.

\section{Traveling-wave-resonator scheme}\label{appB}

\begin{figure}[ptb]
\renewcommand {\thefigure} {A1}
\centering
\includegraphics[width=7 cm]{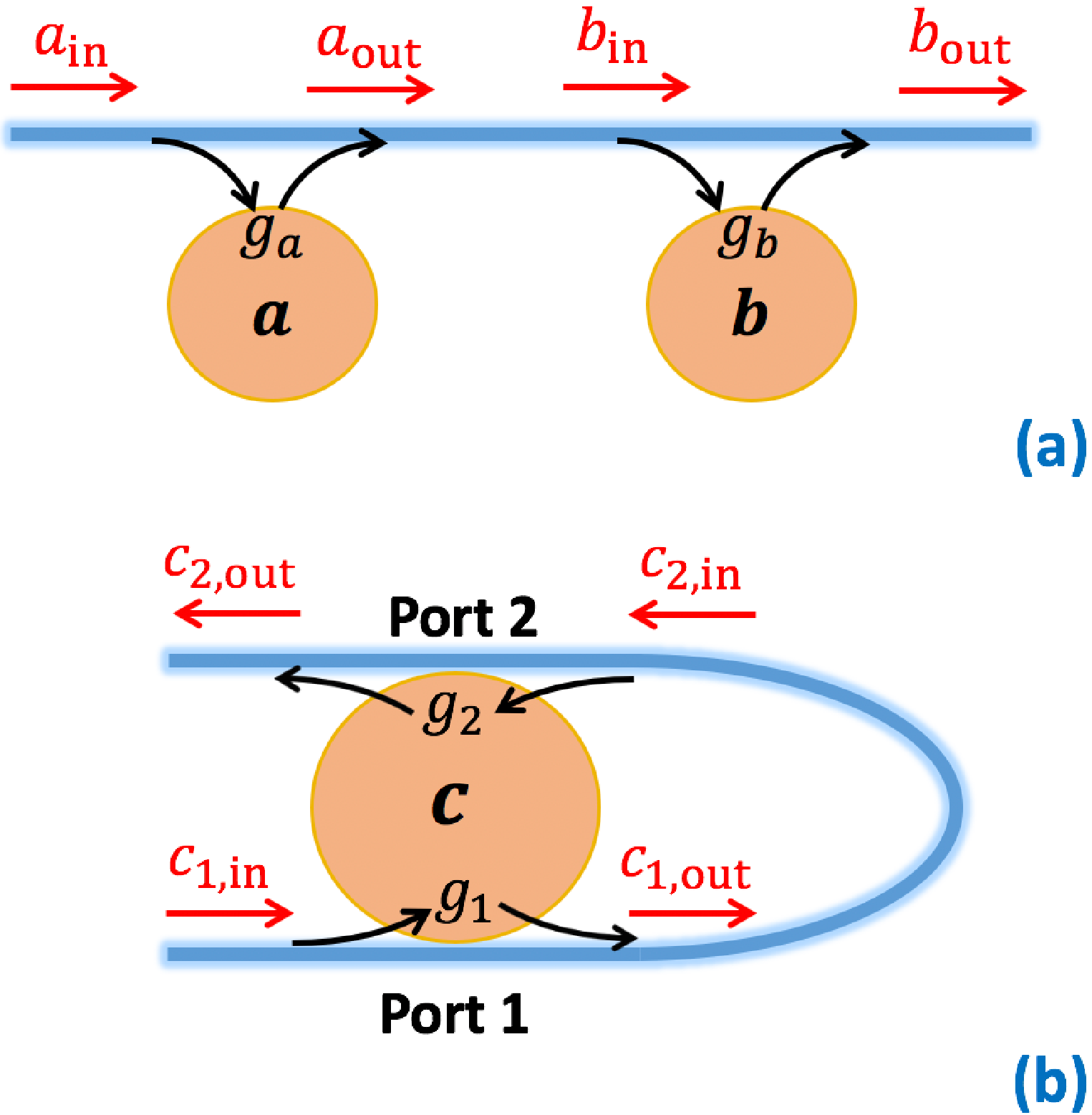}
\caption{Schematic diagrams of (a) two separated traveling-wave resonators side-coupled with a straight waveguide and (b) a single traveling-wave resonator side-coupled with a bent waveguide at two different ports.}\label{figa1}
\end{figure}

If we consider two separated traveling-wave resonators (such as WGM resonators) coupled with a common waveguide, as shown in Fig.~\ref{figa1}(a). The dynamic equations of the two counter-clockwise (CCW) resonator modes $a$ and $b$ can be written as 
\begin{equation}
\begin{split}
&\frac{da}{dt}=-(i\omega_{a}+\kappa_{a})a+\sqrt{\kappa_{a,e}}a_{\textrm{in}},\\
&\frac{db}{dt}=-(i\omega_{b}+\kappa_{b})b+\sqrt{\kappa_{b,e}}b_{\textrm{in}},
\label{B1}
\end{split}
\end{equation}
where $\kappa_{j}=\kappa_{j,i}+\kappa_{j,e}/2$ ($j=a,\,b$) in this case. Here $\omega_{j}$, $\kappa_{j,i}$ and $\kappa_{j,e}$ have the same meaning as those in Eq.~(\ref{eq2}). The external decay rates are not doubled in this case even if the resonators are two-sided due to the directionality of traveling-wave resonator modes. For the same reason, the CCW modes can only be excited by the right-going input field $a_{\textrm{in}}$ coming from the left side of the waveguide. As shown in Fig.~\ref{figa1}(a), the input-output relations of each CCW mode can be written as   
\begin{equation}
\begin{split}
&a_{\textrm{out}}=a_{\textrm{in}}-\sqrt{\kappa_{a,e}}a,\\
&b_{\textrm{in}}=a_{\textrm{out}}e^{i\phi},\\
&b_{\textrm{out}}=b_{\textrm{in}}-\sqrt{\kappa_{b,e}}b
\end{split}
\label{B2}
\end{equation}
in this case, which leads to 
\begin{equation}
\begin{split}
\frac{da}{dt}=&-(i\omega_{a}+\kappa_{a})a+\sqrt{\kappa_{a,e}}a_{\textrm{in}},\\
\frac{db}{dt}=&-(i\omega_{b}+\kappa_{b})b-\sqrt{\kappa_{a,e}\kappa_{b,e}}e^{i\phi}a\\
&+\sqrt{\kappa_{b,e}}e^{i\phi}a_{\textrm{in}}.
\label{B3}
\end{split}
\end{equation}
Eq.~(\ref{B3}) shows that the effective coupling between $a$ and $b$ is unidirectional in this case, i.e., the dynamics of $b$ depends on $a$ but not vice versa. This type of interactions have been used to achieve chiral exceptional points (EPs) in an indirectly coupled WGM resonator system where backscattering induced couplings are also considered~\cite{chiralEP}.   

However, if we consider a single traveling-wave resonator coupled with a bent waveguide at two different ports, as shown in Fig.~\ref{figa1}(b), the dynamic equation of the CCW mode $c$ is given by
\begin{equation}
\frac{dc}{dt}=-(i\omega_{c}+\kappa_{c})c+\sqrt{\kappa_{1,e}}c_{1,\textrm{in}}+\sqrt{\kappa_{2,e}}c_{2,\textrm{in}},
\label{B4}
\end{equation}
where $\kappa_{c}=\kappa_{c,i}+(\kappa_{1,e}+\kappa_{2,e})/2$. $\omega_{c}$, $\kappa_{c,i}$ and $\kappa_{j,e}$ ($j=1,\,2$) have the same meaning as those in Eq.~(\ref{eq5}). As discussed above, we only consider the two forward (the direction from port $1$ to port $2$) input fields $c_{1,\textrm{in}}$ and $c_{2,\textrm{in}}$ at the two ports in this case. With a similar relation shown in Eq.~(\ref{B2}), one can obtain the effective dynamic equation
\begin{equation}
\begin{split}
\frac{dc}{dt}=&-(i\Delta_{c}+\kappa_{c})c-\sqrt{\kappa_{1,e}\kappa_{2,e}}e^{i\phi}c\\
&+(\sqrt{\kappa_{1,e}}+\sqrt{\kappa_{2,e}}e^{i\phi})c_{1,\textrm{in}}
\end{split}
\label{B5}
\end{equation}
and the total input-output relation
\begin{equation}
c_{\textrm{out}}=(c_{1,\textrm{in}}-\sqrt{\kappa_{1,e}}c)e^{i\phi}-\sqrt{\kappa_{2,e}}c.
\label{B6}
\end{equation}
Once again, Eqs.~(\ref{B5}) and (\ref{B6}) can be simplified as
\begin{equation}
\begin{split}
&\frac{dc}{dt}=-[i\Delta_{c}+\kappa_{c,i}+\kappa_{e}(1+e^{i\phi})]c+\sqrt{\kappa_{e}}(1+e^{i\phi})c_{1,\textrm{in}},\\
&c_{\textrm{out}}=c_{1,\textrm{in}}e^{i\phi}-\sqrt{\kappa_{e}}(1+e^{i\phi})c
\end{split}
\label{B7}
\end{equation}
in the case of $\kappa_{1,e}=\kappa_{2,e}=\kappa_{e}$. Now we can find that although the dynamic equations show significant difference for the cases of Fig.~\ref{fig1}(a) and Fig.~\ref{figa1}(a), the self-interference terms appear in the similar form for a single standing-wave or traveling-wave resonator coupled with a waveguide at separated ports. Compared with the standing-wave scheme in Fig.~\ref{fig1}(b), the single-traveling-wave-resonator scheme here shows the advantage in preventing reflections.

\section*{Acknowledgments}

This work was supported by the Science Challenge Project (Grant No. TZ2018003), the National Key R\&D Program of China (Grant No. 2016YFA0301200), the Educational Commission of Jilin Province of China (Grant No. JJKH20190266KJ), and the National Natural Science Foundation of China (Grants No. 11774024, No. 11875011, No. 12074030, and No. U1930402).


\begin{thebibliography}{99}

\bibitem {wQED1} D.~Roy, C.~M. Wilson, and O.~Firstenberg, Colloquium: Strongly interacting photons in one-dimensional continuum, Rev. Mod. Phys. \textbf{89}, 021001 (2017).

\bibitem {wQED2} X.~Gu, A.~F.~Kockum, A.~Miranowicz, Y.-X. Liu, and F.~Nori, Microwave photonics with superconducting quantum circuits, Phys. Rep. \textbf{718}, 1-102 (2017).

\bibitem {circuitS1} O.~Astafiev, A.~M. Zagoskin, A.~Abdumalikov, Y.~A. Pashkin, T.~Yamamoto, K.~Inomata, Y.~Nakamura, and J.~Tsai, Resonance fluorescence of a single artificial atom, Science \textbf{327}, 840-843 (2010).

\bibitem {circuitS2} A.~F. van Lool, A.~Fedorov, K.~Lalumi\`{e}re, B.~C. Sanders, A.~Blais, and A.~Wallraff, Photon-mediated interactions between distant artificial atoms, Science \textbf{342}, 1494-1496 (2013).

\bibitem {circuitNP} Y.~Liu and A.~A. Houck, Quantum electrodynamics near a photonic bandgap, Nat. Phys. \textbf{13}, 48-52 (2017).

\bibitem {optgd1} S.~Faez, P.~T\"{u}rschmann, H.~R. Haakh, S.~G\"{o}tzinger, and V.~Sandoghdar, Coherent interaction of light and single molecules in a dielectric nanoguide, Phys. Rev. Lett. \textbf{113}, 213601 (2014).

\bibitem {optgd2} E.~Vetsch, D.~Reitz, G.~Sagu\'{e}, R.~Schmidt, S.~T. Dawkins, and A.~Rauschenbeutel, Optical interface created by laser-cooled atoms trapped in the evanescent field surrounding an optical nanofiber, Phys. Rev. Lett. \textbf{104}, 203603 (2010).

\bibitem {array1} M.~J. Hartmann, F.~G.~S.~L. Brand\~ao, and M.~B. Plenio, Strongly interacting polaritons in coupled arrays of cavities, Nat. Phys. \textbf{2}, 849-855 (2006).

\bibitem {array2} M.~Notomi, E.~Kuramochi, and T.~Tanabe, Large-scale arrays of ultrahigh-Q coupled nanocavities, Nat. Photonics \textbf{2}, 741-747 (2008).

\bibitem {chiralcp1} R.~Mitsch, C.~Sayrin, B.~Albrecht, P.~Schneeweiss, and A.~Rauschenbeutel, Quantum state-controlled directional spontaneous emission of photons into a nanophotonic waveguide, Nat. Commun. \textbf{5}, 1-5 (2014).

\bibitem {chiralcp2} C.~Sayrin, C.~Junge, R.~Mitsch, B.~Albrecht, D.~O'Shea, P.~Schneeweiss, J.~Volz, and A.~Rauschenbeutel, Nanophotonic optical isolator controlled by the internal state of cold atoms, Phys. Rev. X \textbf{5}, 041036 (2015).

\bibitem {pt1} M.~Fitzpatrick, N.~M. Sundaresan, A.~C.~Y. Li, J.~Koch, and A.~A. Houck, Observation of a dissipative phase transition in a one-dimensional circuit QED lattice, Phys. Rev. X \textbf{7}, 011016 (2017).

\bibitem {pt2} L.~Qiao, Y.-J. Song, and C.-P. Sun, Quantum phase transition and interference trapping of populations in a coupled-resonator waveguide, Phys. Rev. A \textbf{100}, 013825 (2019).

\bibitem {tuqo} M.~Bello, G.~Platero, J.~I. Cirac, and A.~Gonz\'{a}lez-Tudela, Unconventional quantum optics in topological waveguide QED, Sci. Adv. \textbf{5}, eaaw0297 (2019).

\bibitem {nr1} M.-T. Cheng, X.~Ma, J.-W. Fan, J.~Xu, and C.~Zhu, Controllable single-photon nonreciprocal propagation between two waveguides chirally coupled to a quantum emitter, Opt. Lett. \textbf{42}, 2914-2917 (2017).

\bibitem {nr2} W.-B. Yan, W.-Y. Ni, J.~Zhang, F.-Y. Zhang, and H.~Fan, Tunable single-photon diode by chiral quantum physics, Phys. Rev. A \textbf{98}, 043852 (2018).

\bibitem {nr3} Z.~Wang, L.~Du, Y.~Li, and Y.-X. Liu, Phase-controlled single-photon nonreciprocal transmission in a one-dimensional waveguide, Phys. Rev. A \textbf{100}, 053809 (2019).

\bibitem {fan1999} S.~Fan, P.~R. Villeneuve, J.~D. Joannopoulos, M.~J. Khan, C.~Manolatou, and H.~A. Haus, Theoretical analysis of channel drop tunneling processes, Phys. Rev. B \textbf{59}, 15882 (1999).

\bibitem {IEEEo} C.~Manolatou, M.~J. Khan, S.~Fan, P.~R. Villeneuve, H.~A. Haus, and J.~D. Joannopoulos, Coupling of modes analysis of resonant channel add-drop filters, IEEE J. Quantum Electron. \textbf{35}, 1322-1331 (1999).

\bibitem {xiao2008} Y.-F. Xiao, V.~Gaddam, and L.~Yang, Coupled optical microcavities: an enhanced refractometric sensing configuration, Opt. Express \textbf{16}, 12538-12543 (2008).

\bibitem {xiao2010} Y.-F. Xiao, M.~Li, Y.-C. Liu, Y.~Li, X.~Sun, and Q.~Gong, Asymmetric Fano resonance analysis in indirectly coupled microresonators, Phys. Rev. A \textbf{82}, 065804 (2010).

\bibitem {jhli} J.~Li, R.~Yu, C.~Ding, and Y.~Wu, PT-symmetry-induced evolution of sharp asymmetric line shapes and high-sensitivity refractive index sensors in a three-cavity array, Phys. Rev. A \textbf{93}, 023814 (2016).

\bibitem {WongPRL} X.~Yang, M.~Yu, D.-L. Kwong, and C.~W. Wong, All-optical analog to electromagnetically induced transparency in multiple coupled photonic crystal cavities, Phys. Rev. Lett. \textbf{102}, 173902 (2009).

\bibitem {fan2010} J.~Pan, S.~Sandhu, Y.~Huo, M.~Povinelli, J.~S. Harris, M.~M. Fejer, and S.~Fan, Experimental demonstration of an all-optical analog to the superradiance effect in an on-chip photonic crystal resonator system, Phys. Rev. B \textbf{81}, 041101 (2010).

\bibitem {BBLi} B.-B. Li, Y.-F. Xiao, C.-L. Zou, X.-F. Jiang, Y.-C. Liu, F.-W. Sun, Y.~Li, and Q.~Gong, Experimental controlling of Fano resonance in indirectly coupled whispering-gallery microresonators, Appl. Phys. Lett. \textbf{100}, 021108 (2012).

\bibitem {cm1} H.~Huebl, C.~W. Zollitsch, J.~Lotze, F.~Hocke, M.~Greifenstein, A.~Marx, R.~Gross, and S.~T.~B. Goennenwein, High cooperativity in coupled microwave resonator ferrimagnetic insulator hybrids, Phys. Rev. Lett. \textbf{111}, 127003 (2013).

\bibitem {cm2} Y.~Tabuchi, S.~Ishino, T.~Ishikawa, R.~Yamazaki, K.~Usami, and Y.~Nakamura, Hybridizing ferromagnetic magnons and microwave photons in the quantum limit, Phys. Rev. Lett. \textbf{113}, 083603 (2014).

\bibitem {cm3} X.~Zhang, C.-L. Zou, L.~Jiang, and H.~X. Tang, Strongly coupled magnons and cavity microwave photons, Phys. Rev. Lett. \textbf{113}, 156401 (2014).

\bibitem {cm4} M.~Goryachev, W.~G. Farr, D.~L. Creedon, Y.~Fan, M.~Kostylev, and M.~E. Tobar, High-cooperativity cavity QED with magnons at microwave frequencies, Phys. Rev. Applied \textbf{2}, 054002 (2014).

\bibitem {cm5} L.~Bai, M.~Harder, Y.~P. Chen, X.~Fan, J.~Q. Xiao, and C.-M. Hu, Spin pumping in electrodynamically coupled magnon-photon systems, Phys. Rev. Lett. \textbf{114}, 227201 (2015).

\bibitem {cm6} Y.~Cao, P.~Yan, H.~Huebl, S.~T.~B. Goennenwein, and G.~E.~W. Bauer, Exchange magnon-polaritons in microwave cavities, Phys. Rev. B \textbf{91}, 094423 (2015).

\bibitem {gradient} X.~Zhang, C.-L. Zou, N.~Zhu, F.~Marquardt, L.~Jiang, and H.~X. Tang, Magnon dark modes and gradient memory, Nat. Commun. \textbf{6}, 1-7 (2015).

\bibitem {logic} J.~W. Rao, S.~Kaur, B.~M. Yao, E.~R.~J. Edwards, Y.~T. Zhao, X.~Fan, D.~Xue, T.~J. Silva, Y.~S. Gui, and C.-M. Hu, Analogue of dynamic Hall effect in cavity magnon polariton system and coherently controlled logic device, Nat. Commun. \textbf{10}, 1-7 (2019).

\bibitem {mblock} Z.-X. Liu, H.~Xiong, and Y.~Wu, Magnon blockade in a hybrid ferromagnet-superconductor quantum system, Phys. Rev. B \textbf{100}, 134421 (2019).

\bibitem {magSATang} X.~Zhang, C.-L. Zou, L.~Jiang, H.~X. Tang, Cavity magnomechanics, Sci. Adv. \textbf{2}, e1501286 (2016).

\bibitem {LAmag} Y.~Yang, J.~W. Rao, Y.~S. Gui, B.~M. Yao, W.~Lu, and C.-M. Hu, Control of the magnon-photon level attraction in a planar cavity, Phys. Rev. Applied \textbf{11}, 054023 (2019).

\bibitem {magnr} Y.-P. Wang, J.~W. Rao, Y.~Yang, P.-C. Xu, Y.~S. Gui, B.~M. Yao, J.~Q. You, and C.-M. Hu, Nonreciprocity and unidirectional invisibility in cavity magnonics, Phys. Rev. Lett. \textbf{123}, 127202 (2019).

\bibitem {magref} J.~W. Rao , Y.~P. Wang, Y. Yang, T. Yu , Y.~S. Gui, X.~L. Fan, D.~S. Xue, and C.-M. Hu, Interactions between a magnon mode and a cavity photon mode mediated by traveling photons, Phys. Rev. B \textbf{101}, 064404 (2020).

\bibitem {disper1} D.~Dai, Highly sensitive digital optical sensor based on cascaded high-Q ring-resonators, Opt. Express \textbf{17}, 23817-23822 (2009).

\bibitem {disper2} J.~Wang and D.~Dai, Highly sensitive Si nanowire-based optical sensor using a Mach-Zehnder interferometer coupled microring, Opt. Lett. \textbf{35}, 4229-4231 (2010).

\bibitem {disper3} O.~A. Marsh, Y.~Xiong, and N.~Y. Winnie, Slot waveguide ring-assisted Mach-Zehnder interferometer for sensing applications, IEEE J. Quantum Electron. \textbf{23}, 440-443 (2017).

\bibitem {PRDong} S.~Wan, R.~Niu, H.-L. Ren, C.-L. Zou, G.-C. Guo, and C.-H. Dong, Experimental demonstration of dissipative sensing in a self-interference microring resonator, Photon. Res. \textbf{6}, 681-685 (2018).

\bibitem {sensJOSAB} H.~Ren, C.-L. Zou, J.~Lu, Z.~Le, Y.~Qin, S.~Guo, and W.~Hu, Dissipative sensing with low detection limit in a self-interference microring resonator, J. Opt. Soc. of Am. B \textbf{36}, 942-951 (2019).

\bibitem {mapp1} J.~Lenz and S.~Edelstein, Magnetic sensors and their applications, IEEE Sens. J. \textbf{6}, 631-649 (2006).

\bibitem {mapp2} A.~Edelstein, Advances in magnetometry, J. Phys.: Condens. Matter \textbf{19}, 165217 (2007).

\bibitem {mapp3} A.~Grosz, M.~J. Haji-Sheikh, and S.~C. Mukhopadhyay, \emph{High Sensitivity Magnetometers} (Springer, Berlin, 2017).

\bibitem {mmeter1} S.~Forstner, S.~Prams, J.~Knittel, E.~D. van Ooijen, J.~D. Swaim, G.~I. Harris, A.~Szorkovszky, W.~P. Bowen, and H.~Rubinsztein-Dunlop, Cavity optomechanical magnetometer, Phys. Rev. Lett. \textbf{108}, 120801 (2012).

\bibitem {mmeter2} G.~Chatzidrosos, A.~Wickenbrock, L.~Bougas, N.~Leefer, T.~Wu, K.~Jensen, Y.~Dumeige, and D.~Budker, Miniature cavity-enhanced diamond magnetometer, Phys. Rev. Applied \textbf{8}, 044019 (2017).

\bibitem {mmeter3} Y.~Cao and P.~Yan, Exceptional magnetic sensitivity of PT-symmetric cavity magnon polaritons, Phys. Rev. B \textbf{99}, 214415 (2019).

\bibitem {mmeter4} M.~F. Colombano, G.~Arregui, F.~Bonell, N.~E. Capuj, E.~Chavez-Angel, A.~Pitanti, S.~O. Valenzuela, C.~M. Sotomayor-Torres, D.~Navarro-Urrios, and 
M.~V. Costache, Ferromagnetic resonance assisted optomechanical magnetometer, Phys. Rev. Lett. \textbf{125}, 147201 (2020).

\bibitem {Solano} K.~Sinha, P.~Meystre, E.~A. Goldschmidt, F.~K. Fatemi, S.~L. Rolston, and P.~Solano, Non-Markovian collective emission from macroscopically separated emitters, Phys. Rev. Lett. \textbf{124}, 043603 (2020).

\bibitem {longhiret1} S.~Longhi, Superradiance paradox in waveguide lattices, Opt. Lett. \textbf{45}, 3297-3300 (2020).

\bibitem {longhiret2} S.~Longhi, Photonic simulation of giant atom decay, Opt. Lett. \textbf{45}, 3017-3020 (2020).

\bibitem {walls} D.~F. Walls and G.~J. Milburn, \emph{Quantum Optics} (Springer-Verlag, Berlin, 1994).

\bibitem {fan2009} J.-T. Shen and S.~Fan, Theory of single-photon transport in a single-mode waveguide. I. Coupling to a cavity containing a two-level atom, Phys. Rev. A \textbf{79}, 023837 (2009).

\bibitem {synphonon} C.~W. Peterson, S.~Kim, J.~T. Bernhard, and G.~Bahl, Synthetic phonons enable nonreciprocal coupling to arbitrary resonator networks, Sci. Adv. \textbf{4}, eaat0232 (2018).

\bibitem {noise1} M.~R. Foreman, W.-L. Jin, and F.~Vollmer, Optimizing detection limits in whispering gallery mode biosensing, Opt. Express \textbf{22}, 5491-5511 (2014).

\bibitem {noise2} X.~Zhou, L.~Zhang, and W.~Pang, Performance and noise analysis of optical microresonator-based biochemical sensors using intensity detection, Opt. Express \textbf{24}, 18197-18208 (2016).

\bibitem {thermo1} M.~L. Gorodetsky and I.~S. Grudinin, Fundamental thermal fluctuations in microspheres, J. Opt. Soc. Am. B \textbf{21}, 697-705 (2004).

\bibitem {thermo2} W.-L. Jin, X.~Yi, Y.~Hu, B.~Li, and Y.~Xiao, Temperature-insensitive detection of low-concentration nanoparticles using a functionalized high-Q microcavity, Appl. Opt. \textbf{52}, 155-161 (2013).

\bibitem {chiralEP} C.~Wang, X.~Jiang, G.~Zhao, M.~Zhang, C.~W. Hsu, B.~Peng, A.~D. Stone, L.~Jiang, and L.~Yang, Electromagnetically induced transparency at a chiral exceptional point, Nat. Phys. \textbf{16}, 334-340 (2020).

\end{thebibliography}
\end{document}